\documentclass[journal]{IEEEtran}
\usepackage{amsmath,amsfonts}
\usepackage{algorithmic}
\usepackage{algorithm}
\usepackage{array}
\usepackage[caption=false,font=normalsize,labelfont=sf,textfont=sf]{subfig}
\usepackage{textcomp}
\usepackage{stfloats}
\usepackage{url}
\usepackage{verbatim}
\usepackage{graphicx}
\usepackage{cite}
\usepackage{booktabs}

\usepackage[colorlinks=true,
            linkcolor=blue,   
            citecolor=blue,  
            urlcolor=black]{hyperref}
\hypersetup{pdfborder={0 0 0}}
\usepackage[normalem]{ulem} %
\usepackage{soul} %
\usepackage{orcidlink}
\usepackage{xcolor} %
\begin{document}

\title{Multiprotocol Wireless Timer Synchronization for IoT Systems}

\author{Ziyao Zhou\,\orcidlink{0009-0007-0671-9947},
        Tiancheng Cao\,\orcidlink{0000-0002-7259-5192},
        Chen Shen\,\orcidlink{0009-0008-4221-2534},
        Jiaqi Zhang\,\orcidlink{0009-0004-1596-2250},
        Yuting Liu\,\orcidlink{0009-0002-2996-5087},
        Hen-Wei Huang\,\orcidlink{0000-0003-1921-8897}%
\thanks{This work was supported by the Nanyang Assistant Professorship, the MOE Tier 1 grant RG71/24, and the MTC MedTech Programmatic Fund M24N9b0125. \textit{(Corresponding author: Tiancheng Cao.)}}%

\thanks{Ziyao Zhou, Tiancheng Cao, Chen Shen, Jiaqi Zhang, and Hen-Wei Huang are with the School of Electrical and Electronic Engineering, Nanyang Technological University, Singapore. Hen-Wei Huang is also affiliated with the Lee Kong Chian School of Medicine, Nanyang Technological University, Singapore. Yuting Liu is with the Glasgow College, University of Electronic Science and Technology of China, Chengdu, Sichuan, China.}
}

\markboth{IEEE Wireless Communications Letters
,~Vol.~XX, No.~X, Month~Year}%
{Shell \MakeLowercase{\textit{et al.}}: A Sample Article Using IEEEtran.cls for IEEE Journals1}


\maketitle
\begin{abstract}

Accurate time synchronization is essential for Internet of Things (IoT) systems, where multiple distributed nodes must share a common time base for coordinated sensing and data fusion. However, conventional synchronization approaches suffer from nondeterministic transmission latency, limited precision, or restricted bidirectional functionality. This paper presents a protocol-independent wireless timer synchronization method that exploits radio timeslots to transmit precisely timestamped beacons in a proprietary radio mode. By decoupling synchronization from upper-layer packet retransmissions and leveraging hardware-timed radio events, the proposed approach significantly reduces scheduling uncertainty and achieves nanosecond-level synchronization accuracy. Comprehensive experiments evaluate the impacts of synchronization frequency, RSSI, BLE connection interval, and throughput on synchronization performance. The results demonstrate that an optimal synchronization frequency of 1000 Hz yields an approximately 20 ns delay in the absence of communication stack activity while maintaining sub-500 ns accuracy under most realistic BLE traffic conditions. Furthermore, larger connection intervals, lower application throughput, and higher RSSI consistently improve synchronization quality by reducing radio resource contention and packet loss. The proposed scheme provides a general and high-precision synchronization solution suitable for resource-constrained IoT systems.

\end{abstract}


\section{Introduction}
Internet of Things (IoT) systems are widely used for distributed sensing and monitoring applications \cite{Khanna2020Internet, Perwej2019The}. In such networks, devices often need to share a common clock to enable synchronized operations and accurate sensor timestamps for data fusion \cite{Depolli2025Offline,Balasubramanian2023Neural}.

\begin{figure}[t]
    \centering
    \includegraphics[width=\columnwidth]{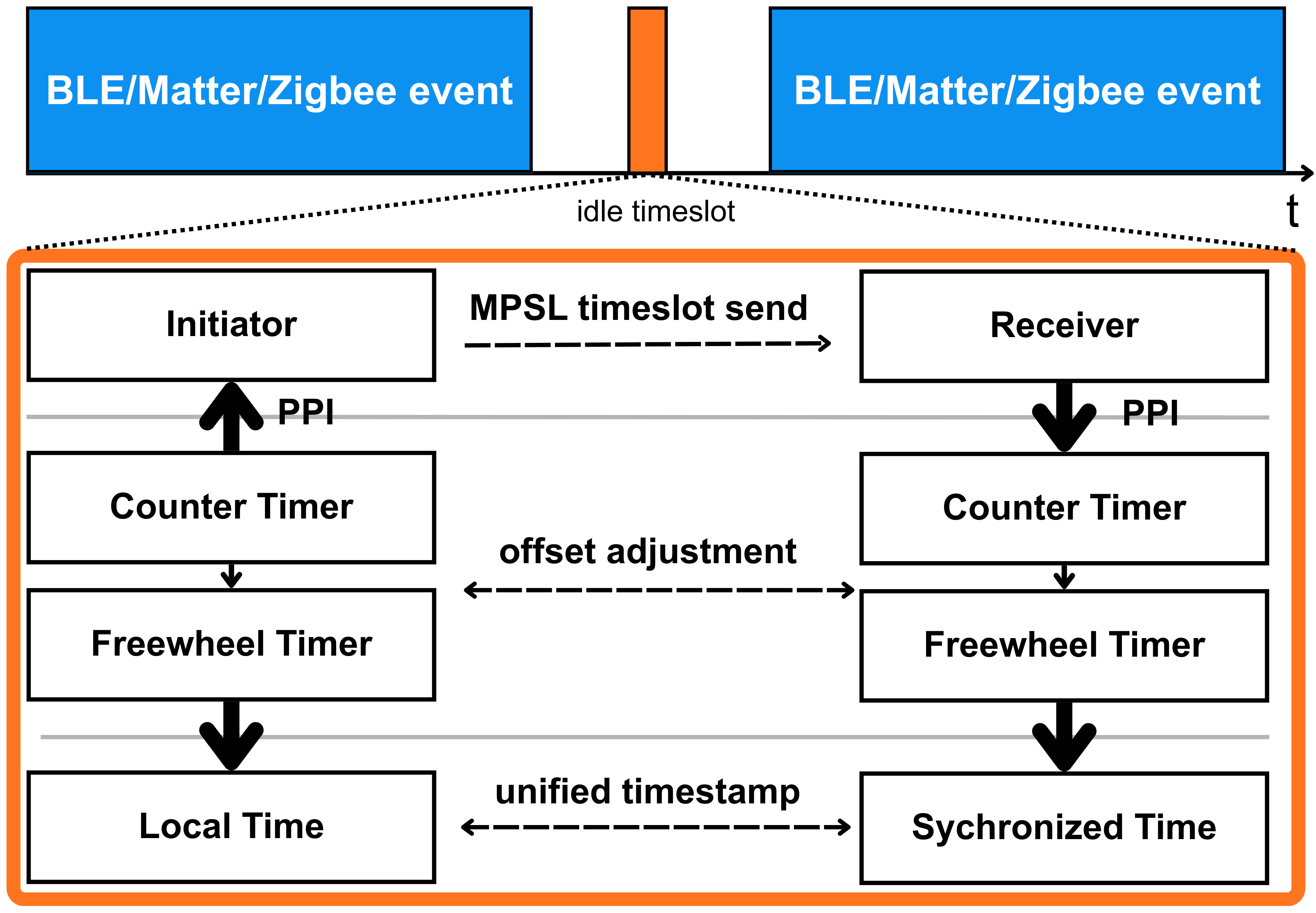}
    \caption{Overview of the proposed wireless timer synchronization mechanism. 
Synchronization beacons are transmitted within reserved multiprotocol service layer (MPSL) timeslots between events. 
Hardware timers and programmable peripheral interconnect (PPI)-based capture enable precise offset estimation and compensation, resulting in a unified global time.}
    \label{fig:system_overview}
\end{figure}

Among the various wireless communication technologies used in IoT systems, Bluetooth Low Energy (BLE) has emerged as one of the most widely deployed protocols due to its low power consumption and extensive ecosystem support \cite{Liu2021A}. As a representative IoT communication technology, BLE has therefore attracted significant attention in the development of wireless time synchronization mechanisms. In BLE-based IoT networks, existing synchronization algorithms can generally be divided into two categories: advertising-based approaches and connection-oriented approaches, each with distinct advantages and trade-offs \cite{Landra2025SharkTooth}.

Advertising-based methods, such as CheepSync and BlueSync, achieve very high accuracy (tens of microseconds to hundreds of nanoseconds) by leveraging low-level timestamping and hardware features \cite{Sridhar2015Cheepsync, Asgarian2022BlueSync}. However, they operate only in advertising mode, which supports only unidirectional communication and cannot enable read, write, or notify operations. Therefore, they are unsuitable for interactive IoT networks.

Connection-oriented approaches enable bidirectional communication. Conventional methods based on connection achieve sub-millisecond accuracy but do not compensate for clock drift and often require repeated reconnections \cite{Dian2017A}. Application-layer regression methods improve deployability and compatibility, but their validation is limited to small-scale setups, and they are sensitive to transmission delays or packet loss \cite{Li2023ApplicationLayer}. Neural-network-based solutions improve robustness to transmission variability, yet their scalability and training transparency remain unclear \cite{Balasubramanian2023Neural}.
Another approach is offline correction, which enhances synchronization robustness but cannot support real-time synchronization \cite{Depolli2025Offline}. To the best of our knowledge, no existing BLE synchronization method simultaneously achieves high precision, bidirectional communication, real-time operation, and scalability to large networks.

This article proposes a simplified yet high-precision synchronization scheme that exploits radio timeslots available between protocol activities to transmit and receive precisely timestamped beacon signals, as illustrated in Fig.~\ref{fig:system_overview}. We first evaluate the proposed method under varying synchronization frequencies and signal quality conditions. Subsequently, using BLE as a representative case, we further examine the impacts of connection interval and data throughput on synchronization performance. Overall, this study presents a high-precision synchronization framework that enables efficient time alignment among distributed IoT devices and is protocol-agnostic, allowing deployment across diverse wireless communication standards such as BLE, Zigbee, and Matter.

\section{Methodology}

We selected a pair of nRF52DK development boards equipped with the nRF52832 chipset. The synchronization firmware was adapted from Nordic Semiconductor’s demo \cite{nordic_auko_nrf5_ble_timesync_demo}. The original example was modified such that one board operated as the synchronization initiator and the other as the synchronization receiver. The modified synchronization procedure is summarized in Algorithm~\ref{alg:sync_core} and Fig.~\ref{fig:system_overview}.

\begin{algorithm}[t]
\caption{hardware-assisted timeslot synchronization}
\label{alg:sync_core}
\begin{algorithmic}[1]

\STATE \textbf{setup:} start timer $T$, counter $C$, request periodic timeslots

\STATE \textbf{TX node:}
\FOR{each granted timeslot}
\STATE capture $(T_i,C_i)$ at radio READY; transmit $\{T_i,C_i\}$
\ENDFOR

\STATE \textbf{RX node:}
\FOR{each granted timeslot}
\IF{beacon detected}
\STATE capture $(T_r,C_r)$
\STATE $\Delta=(C_r-C_i)T_{\max}+T_r-T_i$
\STATE $T\leftarrow T-\Delta$
\ENDIF
\ENDFOR

\end{algorithmic}
\end{algorithm}
Instead of relying on upper-layer packet exchanges, which may introduce retransmissions and nondeterministic delays, the proposed method uses the Nordic MPSL to request dedicated radio timeslots scheduled between normal wireless protocol activities. Within these granted timeslots, the initiator switches the radio to a proprietary mode and transmits precisely timed synchronization beacons containing its current high-resolution timer value.

Each node maintains two hardware timers driven by a certain clock domain. Upon receiving a synchronization beacon, the receiver captures the local timer value at the exact reception instant using hardware-based timestamping. The receiver estimates the clock offset by comparing the received timestamp with its own timestamp. This offset is then compensated through timer adjustment, enabling both devices to converge to a unified time base. 

Because the beacon transmission and timestamp capture are handled using PPI and radio triggers within reserved timeslots, the timing uncertainty introduced by protocol retransmissions, CPU latency, and software scheduling is minimized. Moreover, since the synchronization beacons are exchanged in a proprietary radio mode decoupled from the active wireless stack, the same synchronization principle can operate concurrently with different communication protocols, such as BLE, Zigbee, or Matter. This design enables protocol-independent, high-precision clock alignment while maintaining compatibility with standard wireless communication stacks, without introducing additional communication overhead or reducing application-layer throughput.

Both devices are designed to toggle a designated GPIO pin using hardware timer-triggered events. To evaluate synchronization performance, an oscilloscope was connected simultaneously to the GPIO outputs of both boards. Synchronization accuracy was quantified by measuring the time difference between the rising edges of the two waveforms. To assess the power consumption introduced by the synchronization mechanism, both boards were connected using the Nordic Power Profiler Kit II to sample the current.

In addition, the number of synchronization packets transmitted by the initiator and the number successfully received by the receiver were recorded throughout the experiments. By comparing these counts, we quantified packet reception performance, synchronization robustness, and the influence of wireless channel conditions on the proposed synchronization approach.

\subsection{Relationship Between Synchronization Performance and Synchronization Frequency}

Synchronization performance depends on the transmission frequency of synchronization packets. Because the timeslot scheduler operates with a discrete timing resolution, the requested transmission frequency must be aligned to this granularity, resulting in a quantized timeslot spacing. In practice, the effective rate may deviate slightly due to timeslot blocking, cancellation, or scheduling jitter.

To investigate the impact of synchronization frequency on synchronization performance, the original implementation was evaluated without any wireless communication protocol stack running concurrently. The two boards were placed in close proximity to minimize channel attenuation and external interference. The requested transmission frequency was configured from 1~Hz to 3125~Hz. For each required frequency setting, synchronization accuracy was recorded, together with the actual number of synchronization packets transmitted by the initiator and successfully received by the receiver. Meanwhile, the energy consumption of both the transmitting and receiving nodes was measured throughout the experiments.

\subsection{Relationship Between Synchronization Performance and Signal Quality}

In addition to transmission frequency, signal quality is also expected to influence synchronization performance. When synchronization packets are transmitted under poor channel conditions and cannot be reliably demodulated at the receiver, packet loss increases and synchronization accuracy degrades. In this study, RSSI was used as a quantitative metric to characterize signal quality.

The initiator and receiver were both positioned in the air. By progressively increasing the separation distance, the direct path loss between the transmitter and receiver was intentionally intensified, resulting in a gradual degradation of the RSSI. During the experiments, synchronization accuracy, the number of transmitted and successfully received synchronization packets, and the power consumption of both devices were recorded.

Subsequently, a BLE-based implementation was employed, in which the peripheral device was positioned at the corresponding separation distances, while the central device remained fixed at the same location to record the associated RSSI values. In both experiments, the transmission power of the device was kept constant. By integrating the results from these two measurement sets, the relationship between RSSI and synchronization performance, packet statistics, and power consumption was systematically characterized.
\subsection{Relationship Between Synchronization Performance and BLE Connection Interval}

Using the synchronization algorithm alone is insufficient for practical applications; it must operate together with data acquisition and wireless transmission to fulfill fundamental IoT functionalities. To validate the effectiveness of the proposed synchronization method when operating alongside a communication protocol, we selected BLE as a representative.

In this experiment, the synchronization algorithm was integrated with the BLE stack. One board was configured as a BLE peripheral device and the other as a BLE central device. Either the peripheral or the central could be configured as the synchronization initiator, while the other board functioned as the synchronization receiver.

A key feature of BLE is its duty-cycled communication pattern defined by the connection interval, which enables low-power operation. The connection interval can be configured from 7.5 ms to 4 s \cite{Bulić2019Data}. Different connection intervals may affect the number of radio timeslots that MPSL can successfully obtain, thereby influencing synchronization performance.

To investigate this effect, we swept the connection interval from its minimum to maximum value and recorded the synchronization delay. Notably, since the central and peripheral devices have different roles and scheduling responsibilities in BLE, two configurations were evaluated: (1) the central acting as the synchronization initiator and the peripheral as the receiver, and (2) the peripheral acting as the synchronization initiator and the central as the receiver.

\subsection{Relationship Between Synchronization Performance and BLE Throughput}

For IoT devices, throughput is another critical consideration. Many common sensors, such as barometers, photoplethysmography (PPG), and electrocardiography (ECG), require relatively low data rates (typically tens of kbps)\cite{You2024An}. Motion capture applications may demand around 100 kbps \cite{8501957}, whereas video transmission can approach the maximum achievable bandwidth. In theory, as the application data rate increases, more packets occupy the transmission queue and connection events, leaving fewer opportunities for MPSL to obtain radio timeslots for synchronization, thereby potentially degrading synchronization performance.

To investigate the impact of throughput on synchronization, the default connection interval (50 ms) was selected for the experiments. The application data transmission rate was varied by controlling the data transmission delay. The effective throughput was swept from 0 kbps to 1200 kbps. For two BLE configurations, the delay between the GPIO signals of the transmitting and receiving nodes was recorded.

\section{Results and Discussion}
\subsection{Relationship Between Synchronization Performance and Synchronization Frequency}

\begin{figure}[t]
    \centering
    \includegraphics[width=\linewidth]{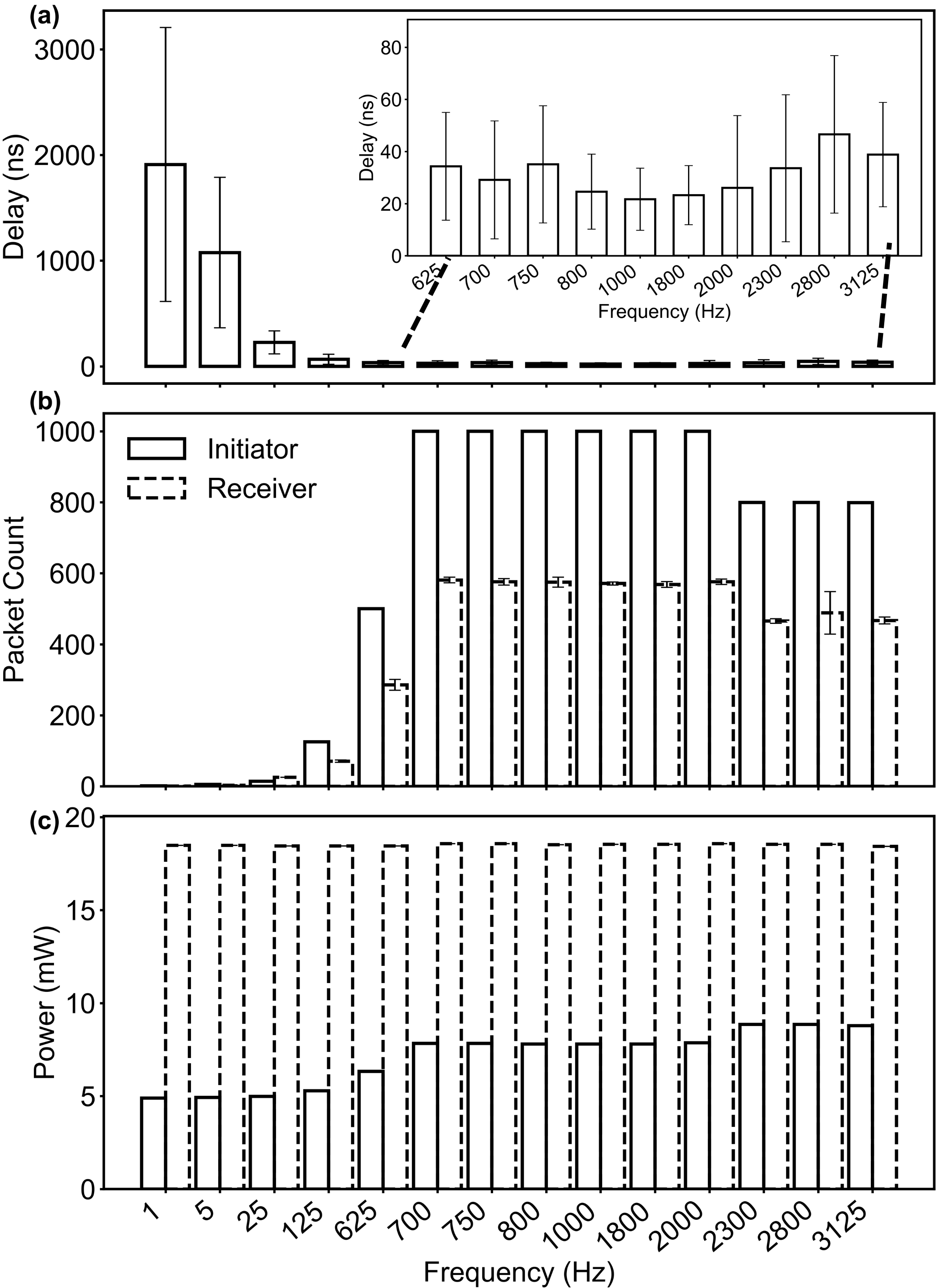}
    \caption{Relationship between synchronization performance and synchronization frequency. (a) Measured delay. (b) Number of packets transmitted by the initiator and received by the receiver. (c) Power consumption of the initiator and receiver.}
    \label{fig:sync_frequency}
\end{figure}

The relationship between synchronization performance and synchronization frequency is shown in Fig. \ref{fig:sync_frequency}. Fig. \ref{fig:sync_frequency}(a) presents the synchronization error between the initiator and the receiver. It can be observed that from 1 Hz to 700 Hz, the overall delay is relatively large and exhibits significant variation. As the frequency increases, synchronization performance improves progressively, and the delay decreases accordingly. In the range of 700–2000 Hz, the synchronization performance remains relatively stable, with the delay reaching an optimal value of approximately 25 ns. However, when the frequency exceeds 2300 Hz, synchronization performance begins to degrade, and the delay increases to approximately 40 ns. These results indicate that synchronization performance is maximized within an optimal frequency range of 700–2000 Hz.

Fig. \ref{fig:sync_frequency}(b) explains why increasing the synchronization frequency does not continuously improve synchronization performance. From 1 Hz to 700 Hz, the requested frequency is positively correlated with the number of transmitted synchronization packets (not strictly proportional). As the frequency increases, more packets are transmitted and received, resulting in progressively improved synchronization performance. However, in the range of 750–2000 Hz, the number of transmitted synchronization packets saturates at approximately 1000 packets per second and no longer increases. When the requested frequency exceeds 2300 Hz, the number of transmitted packets decreases to around 800 per second. These results indicate that a synchronization frequency of approximately 1000 Hz effectively achieves optimal 1000 transmitted packets per second, whereas excessively high requested frequencies become counterproductive and reduce the actual number of successful synchronization transmissions.

Fig. \ref{fig:sync_frequency}(c) illustrates the power consumption of the synchronization initiator and receiver as a function of frequency. Overall, the initiator exhibits relatively low power consumption, whereas the receiver consumes significantly more power. This difference arises because the initiator operates in a short, scheduled transmission mode, while the receiver continuously listens for incoming synchronization packets. As the frequency increases, the initiator’s power consumption gradually rises, whereas the receiver’s power remains nearly constant. 

\vspace{-6pt}

\subsection{Relationship Between Synchronization Performance and Signal Quality}

\begin{figure}[t]
    \centering
    \includegraphics[width=\linewidth]{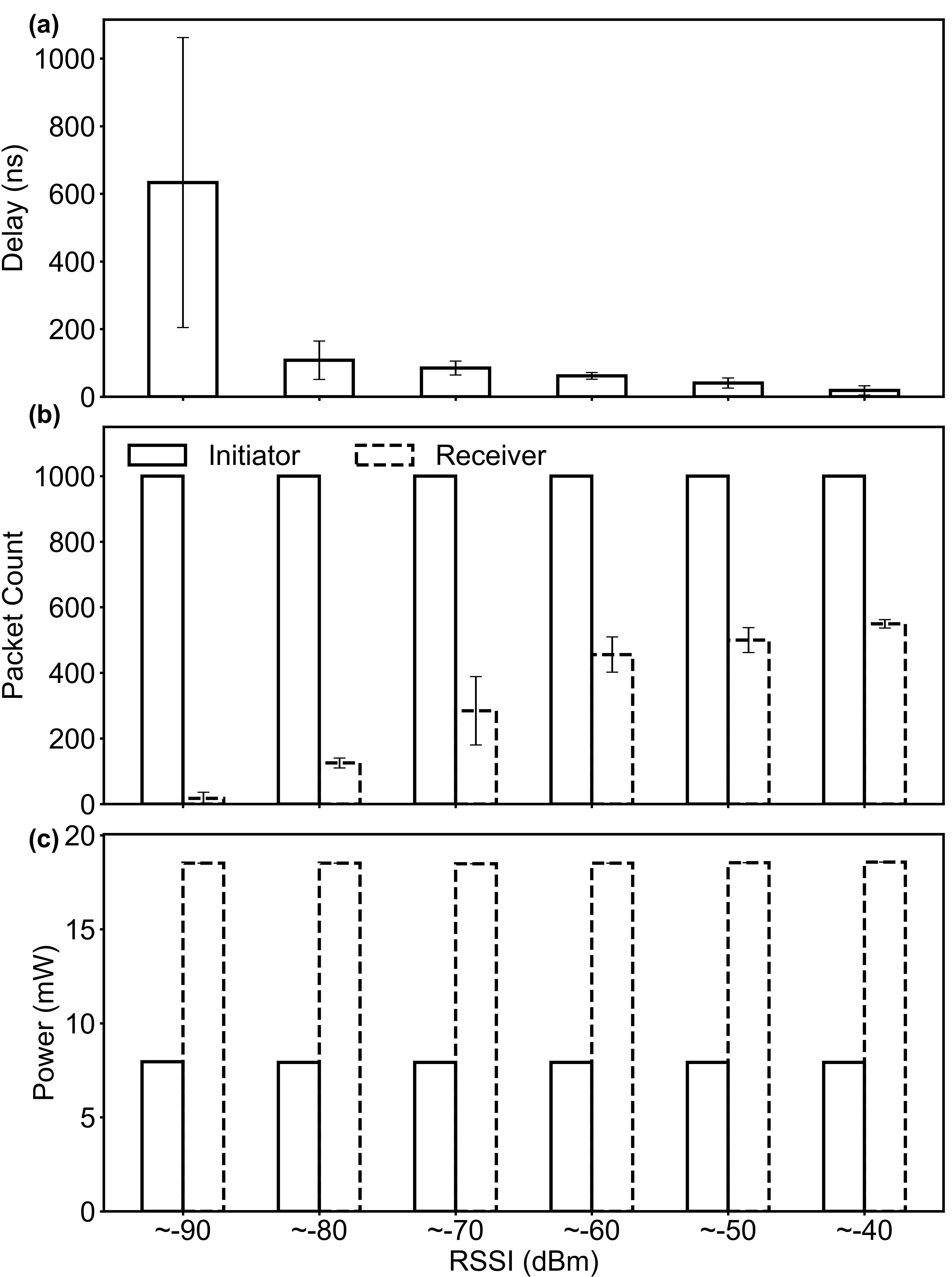}
    \caption{Relationship between synchronization performance and signal quality (RSSI). (a) Measured synchronization delay. (b) Number of synchronization packets transmitted by the initiator and received by the receiver. (c) Power consumption of the initiator and receiver.}
    \label{fig:sync_rssi}
\end{figure}

Using the optimal synchronization frequency of 1000 Hz, the relationship between synchronization performance and signal quality (RSSI) is shown in Fig. \ref{fig:sync_rssi}. Fig. \ref{fig:sync_rssi}(a) presents the synchronization delay under different RSSI levels. As RSSI increases, the delay decreases significantly, from 633.4 ns at approximately -80 dBm to 18.6 ns at approximately -40 dBm. Low RSSI significantly increases delay and variability because weak signals reduce the number of successfully demodulated synchronization packets.

This observation is further confirmed in Fig. \ref{fig:sync_rssi}(b). Although the initiator consistently transmits 1000 synchronization packets per second, the receiver can decode only about 17 packets per second at an RSSI of approximately -80 dBm. In contrast, when the RSSI improves to approximately -40 dBm, around 550 packets per second are successfully received, leading to significantly improved synchronization accuracy.

As shown in Fig. \ref{fig:sync_rssi}(c), the power consumption of both the initiator and the receiver remains nearly constant throughout the RSSI sweep. These results indicate that, when using the proposed synchronization algorithm, maintaining sufficient signal quality is critical for achieving high synchronization accuracy. In practical IoT deployments, devices should either be placed in close proximity or operate with enhanced radio configurations (e.g., higher transmit power or RF amplification) to preserve adequate RSSI and ensure reliable synchronization performance.

\subsection{Relationship Between Synchronization Performance and BLE Connection Interval}

\begin{figure}[t]
    \centering
    \includegraphics[width=\linewidth]{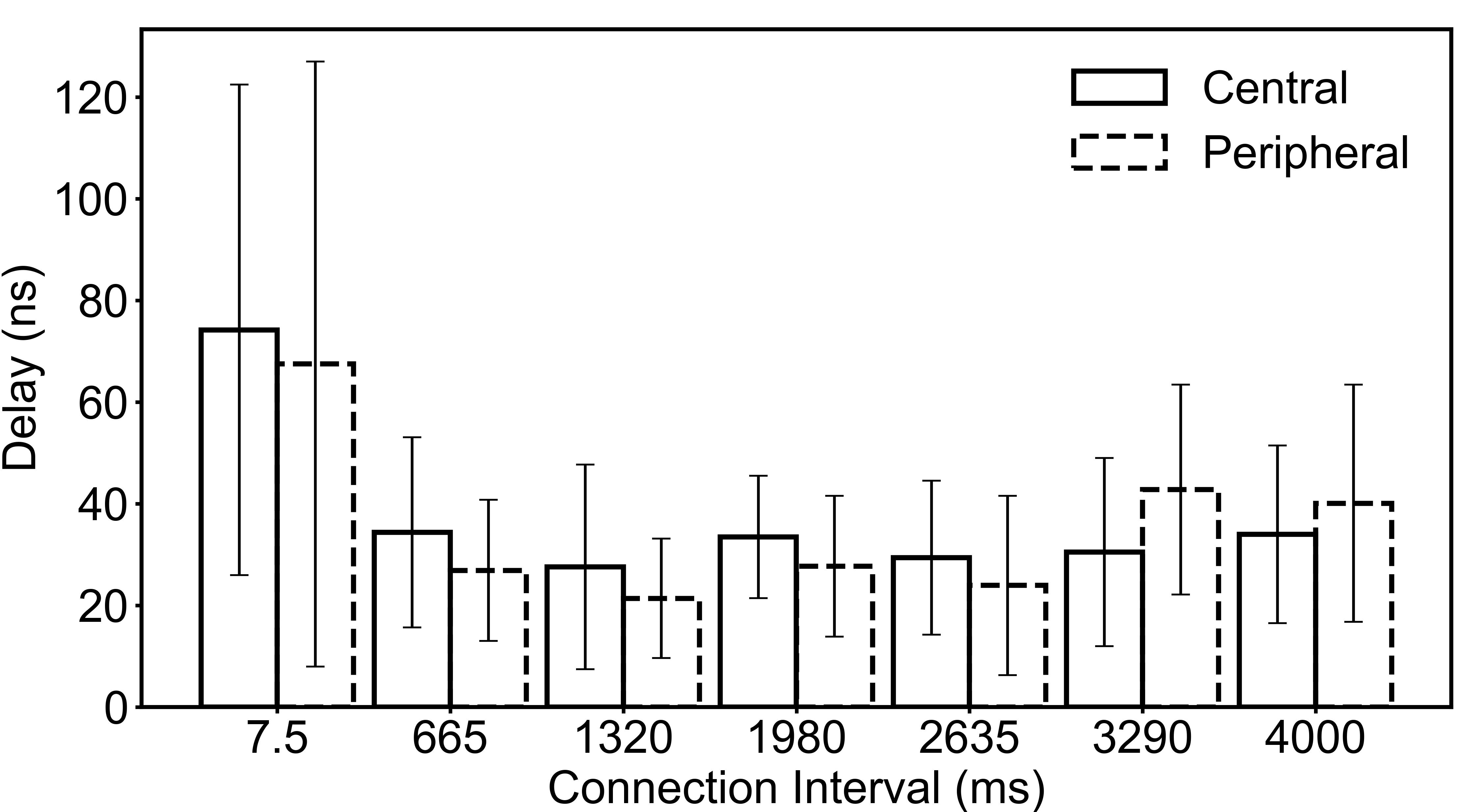}
    \caption{Relationship between synchronization performance and BLE connection interval.}
    \label{fig:sync_conn_interval}
\end{figure}

Two boards were configured as a BLE peripheral and central device to validate the effectiveness of synchronization when integrated with the BLE. By varying the connection interval and alternating the synchronization initiator between the central and the peripheral, the resulting synchronization delays were measured, as shown in Fig. \ref{fig:sync_conn_interval}.

It shows that the synchronization delay generally decreases as the connection interval increases, reducing from 68.8~ns at a 7.5~ms interval to 40.1~ns at a 4000~ms interval. This trend can be attributed to reduced radio resource contention at longer connection intervals. When the interval is short, frequent BLE intervals occupy the radio more intensively, increasing the likelihood of timeslot blocking, scheduling shifts, and timing jitter. As the interval increases, longer idle periods become available, allowing synchronization timeslots to be scheduled with less interference. This improved temporal isolation reduces scheduling uncertainty and leads to smaller synchronization delay.

Furthermore, the synchronization delay and its variation are nearly identical whether the central or the peripheral acts as the synchronization initiator. This is because the proposed method operates at the MPSL timeslot level and is independent of the BLE logical role. 

\subsection{Relationship Between Synchronization Performance and BLE Throughput}

\begin{figure}[t]
    \centering
    \includegraphics[width=\linewidth]{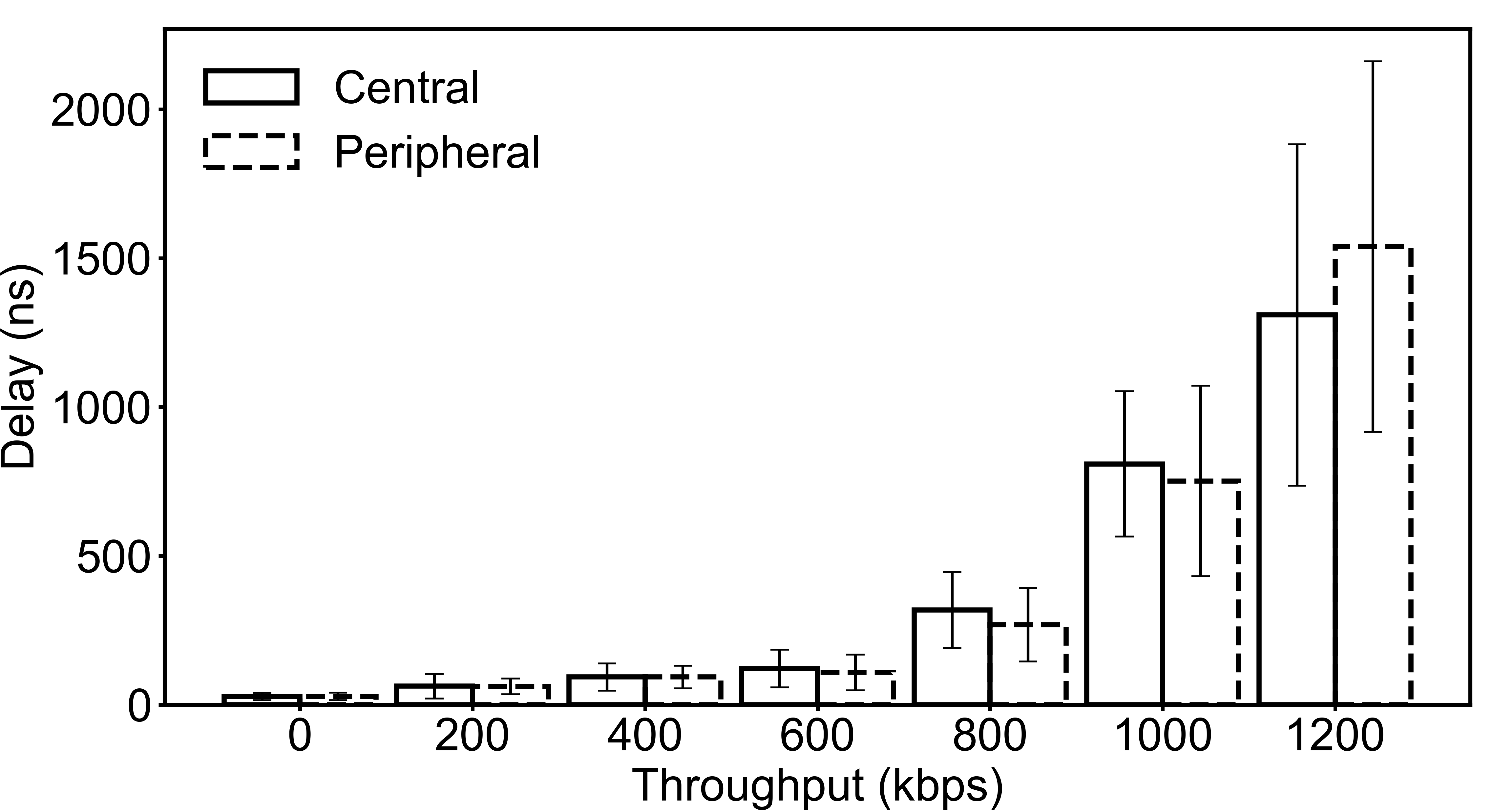}
    \caption{Relationship between synchronization performance and BLE throughput.}
    \label{fig:sync_throughput}
\end{figure}

Fig. \ref{fig:sync_throughput} illustrates the impact of throughput on synchronization performance. The throughput was scanned from 0 to 1200~kbps. It can be observed that higher throughput leads to degraded synchronization performance. When the throughput is 0~kbps, the delay is approximately 30~ns, whereas it increases sharply to around 1500~ns at 1200~kbps. 

As throughput increases, BLE connection events occupy more radio resources, which increases timeslot blocking and scheduling jitter, thereby enlarging the synchronization delay. Nevertheless, since most BLE-based IoT applications require only moderate throughput, high synchronization accuracy can still be maintained in practical scenarios. In addition, although synchronization performance is largely independent of whether the central or the peripheral acts as the initiator, assigning the peripheral this role can be advantageous in BLE-based IoT systems because peripheral devices are typically constrained in computational capability and energy resources \cite{11222604}.

\begin{table}[t]
\caption{Comparison of synchronization methods.}
\label{tab:sync_comparison}
\centering
\small
\begin{tabular}{p{2.8cm} p{3cm} p{2cm}}
\hline
\textbf{Method} & \textbf{Mechanism} & \textbf{Accuracy} \\
\hline
\textbf{This work} & \textbf{Protocol-agnostic} & $\mathbf{\approx 22\,\mathrm{ns}}$ \\

CheepSync \cite{Sridhar2015Cheepsync} 
& Bluetooth GAP 
& $10\,\mathrm{\mu s}$ \\

BlueSync \cite{Asgarian2022BlueSync} 
& Bluetooth GAP 
& $320\,\mathrm{ns}$ \\

\cite{Dian2017A} 
& Bluetooth GATT 
& $750\,\mathrm{\mu s}$ \\

\cite{Balasubramanian2023Neural}
& Neural Network
& $\approx 5\,\mathrm{ms}$ \\

\cite{Depolli2025Offline}
& Offline
& $\approx 1\,\mathrm{ms}$ \\

\hline
\end{tabular}
\end{table}

As summarized in Table \ref{tab:sync_comparison}, the proposed method outperforms existing BLE synchronization approaches by achieving nanosecond-level precision, whereas conventional techniques typically exhibit microsecond- or millisecond-level errors. In addition, its protocol-agnostic design enables straightforward extension to diverse wireless standards, supporting broader deployment in heterogeneous IoT systems.

\section{Conclusion}

This work presents a protocol-agnostic wireless timer synchronization framework for IoT systems. By utilizing multiprotocol timeslots and proprietary radio beacons, the proposed method achieves nanosecond-level synchronization accuracy. Experimental evaluations demonstrate that synchronization performance depends on synchronization frequency, RSSI, BLE connection interval, and throughput. An optimal synchronization frequency range was identified, and sufficient RSSI was shown to be critical for maintaining precision. Furthermore, the relatively long connection interval and lower throughput of BLE can also be beneficial for synchronization performance.

Future work will focus on large-scale multi-node network validation. In addition, integrating the proposed method into devices from a broader range of manufacturers, as well as emerging low-power wireless standards, and evaluating its performance in fully embedded IoT platforms will further demonstrate its practical applicability.%

\bibliographystyle{IEEEtran}
\bibliography{main}

 \vspace{-43pt}

\end{document}